\documentclass[preprint,12pt, a4paper]{elsarticle}
\usepackage{amssymb}
\usepackage{subcaption}
\usepackage{wrapfig}
\usepackage{amsmath}
\usepackage{graphicx}
\usepackage{xcolor}
\usepackage{booktabs}
\usepackage{colortbl}
\usepackage{hyperref}
\usepackage[nameinlink]{cleveref}
\journal{SoftwareX}

\begin{document}
\renewcommand{\labelenumii}
{\arabic{enumi}.\arabic{enumii}}

\begin{frontmatter}

\title{PyPOD-GP: Using PyTorch for Accelerated Chip-Level Thermal Simulation of the GPU}

\author[label1]{Neil He}
\ead{neil.he@yale.edu}
\author[label2]{Ming-Cheng Cheng}
\ead{mcheng@clarkson.edu}
\author[label2]{Yu Liu\corref{cor1}}
\address[label1]{Department of Mathematics, Yale University}
\address[label2]{Department of Electrical and Computer Engineering, Clarkson University}
\ead{yuliu@clarkson.edu}
\cortext[cor1]{Corresponding Author}
\begin{abstract}
The rising demand for high-performance computing (HPC) has made full-chip dynamic thermal simulation in many-core GPUs critical for optimizing performance and extending device lifespans.
Proper orthogonal decomposition (POD) with Galerkin projection (GP) has shown to offer high accuracy and massive runtime improvements over direct numerical simulation (DNS).
However, previous implementations of POD-GP use MPI-based libraries like PETSc and FEniCS and face significant runtime bottlenecks.
We propose a \textbf{Py}Torch-based \textbf{POD-GP} library (PyPOD-GP), a GPU-optimized library for chip-level thermal simulation.
PyPOD-GP achieves over $23.4\times$ speedup in training and over $10\times$ speedup in inference on a GPU with over 13,000 cores, with just $1.2\%$ error over the device layer.

\end{abstract}

\begin{keyword}
GPU Thermal Simulation \sep Proper Orthogonal Decomposition (POD) \sep Finite Element Method \sep PyTorch
\end{keyword}

\end{frontmatter}
\begin{table}[!h]
\begin{tabular}{|l|p{6.5cm}|p{6.5cm}|}
\hline
\textbf{Nr.} & \textbf{Code metadata description} & \textbf{Metadata} \\
\hline
C1 & Current code version & v1.0 \\
\hline
C2 & Permanent link to code/repository used for this code version &\url{https://github.com/784956494/PyPOD-GP} \\
\hline
C3  & None\\
\hline
C4 & Legal Code License   & GLP \\
\hline
C5 & Code versioning system used & git\\
\hline
C6 & Software code languages, tools, and services used & Python \\
\hline
C7 & Compilation requirements, operating environments \& dependencies & Python, PyTorch, numpy, pandas, h5py, Dolfin\\
\hline
C8 & If available Link to developer documentation/manual &\url{https://github.com/784956494/PyPOD-GP/blob/main/README.md} \\
\hline
C9 & Support email for questions & yuliu@clarkson.edu\\
\hline
\end{tabular}
\caption{Code metadata}
\label{codeMetadata} 
\end{table}

\section{Motivation and significance}
Advancements in chip design have significantly increased processor power density~\cite{guggari2021analysis}, leading to high temperature gradients and hot spots that degrade performance and reliability~\cite{heinig2014thermal}. Dynamic thermal management has been implemented to mitigate these issues~\cite{silva2021dynamic, mohammed2019temperature}, showing impressive performance gains in heterogeneous multi-core processors~\cite{kim2020adaptive}. However, these thermal management systems still require efficient and highly accurate thermal simulation. Direct numerical simulations (DNS) provide accurate temperature solutions but are computationally expensive due to their high degrees of freedom (DoF). Alternative methods~\cite{huang2007improved, yuang2021pact, ziabari2014power, zhange2017machine, sultan2019survey} improve efficiency but sacrifice accuracy and/or resolution. An approach combining proper orthogonal decomposition (POD) with Galerkin projection (GP)~\cite{berkooz1993pod} enables efficient and accurate simulations~\cite{lin2023podtherm, lin2024ensemble}. The POD-GP method trains a reduced set of ODEs and POD modes, then computes temperature solutions as a linear combination of these modes weighted by ODE solutions. This approach achieves over five orders of magnitude speedup compared to DNS while maintaining high accuracy~\cite{lin2023podtherm, lin2024ensemble}. However, previous implementations use CPU-based FEM libraries like FEniCS and PETSc~\cite{lin2023podtherm, lin2024ensemble} and thus face runtime bottlenecks during training and inference, limiting their practicality for GPUs with many cores. To address this, we propose PyPOD-GP, a GPU-accelerated implementation leveraging PyTorch's tensor operations. Compared to CPU-based implementations, PyPOD-GP achieves over \textbf{23.4$\times$ speedup} in training  and \textbf{10$\times$ speedup} in inference on an NVIDIA Tesla Volta GV100 GPU, a GPU with 13,440 cores. It also maintains an error of only $1.2\%$ on the heating layer, demonstrating PyPOD-GP's potential for efficient and accurate thermal simulation in large-scale GPU architectures.
\section{Software description}
PyPOD-GP is an open source, PyTorch-based Python library for GPU-accelerated chip-level thermal simulation using the POD-GP method and its variants~\cite{lin2023podtherm,lin2024ensemble}. By eliminating the need for MPI-based libraries, PyPOD-GP significantly accelerates high-resolution thermal simulations for many-core processors with high accuracy. The library offers a user-friendly workflow managed through a single Python object, requiring only the path to training data and chip properties, making it accessible for non-programmers. The software, including installation instructions and test scripts for an AMD ATHLON II X4 610e CPU, is available on GitHub~\cite{pypodgp2024github}. The following sections detail the software's functionality and the process for thermal simulation.

\subsection{Software architecture}\label{arch}
\textbf{The POD-GP Method.} 
The library implements the POD-GP method for thermal simulation, summarized here to clarity. Further details are available in~\cite{lin2023podtherm}. The thermal solution is approximated as a linear combination of $d$ POD modes $\eta_i$ as $T(\vec{r}, t) = \sum_{i=1}^d b_i \eta_i(\vec{r})$, where $b_i$ are weighting coefficients. The POD modes maximize the mean square inner product of the thermal solution with each mode and are obtained by solving the eigenvalue problem for the time-steps correlation matrix $A$, whose entries are inner products of thermal data at two time steps. The coefficients $b_i$ solve the ODE given by

\begin{equation}\label{GP_equation}
    \int_\Omega \left(\eta_i \frac{\partial \rho C_s T}{\partial t} + \nabla \eta_i \cdot \kappa \nabla T\right) d\Omega = \int_\Omega \eta_i P_d(\vec{r}, t) d\Omega - \int_S \eta_i (-\kappa \nabla T \cdot \vec{n}) dS,
\end{equation}
where $\kappa, \rho, C_s$ are thermal conductivity, density and specific heat, respectively, $P_d$ is the interior power density, $S$ is the boundary surface and and $\overrightarrow{n}$ is its outward normal vector. The expression in \cref{GP_equation} can be expressed in matrix form as $C\frac{d\vec{b}}{dt} + G\vec{b} = \vec{P}$, referred to as the $C,G$ matrices and $\vec{P}$ vector.

\begin{wrapfigure}{r}{0.50\textwidth} 
\vspace{-20pt}
  \begin{center}    \centerline{\includegraphics[width=0.50\textwidth]{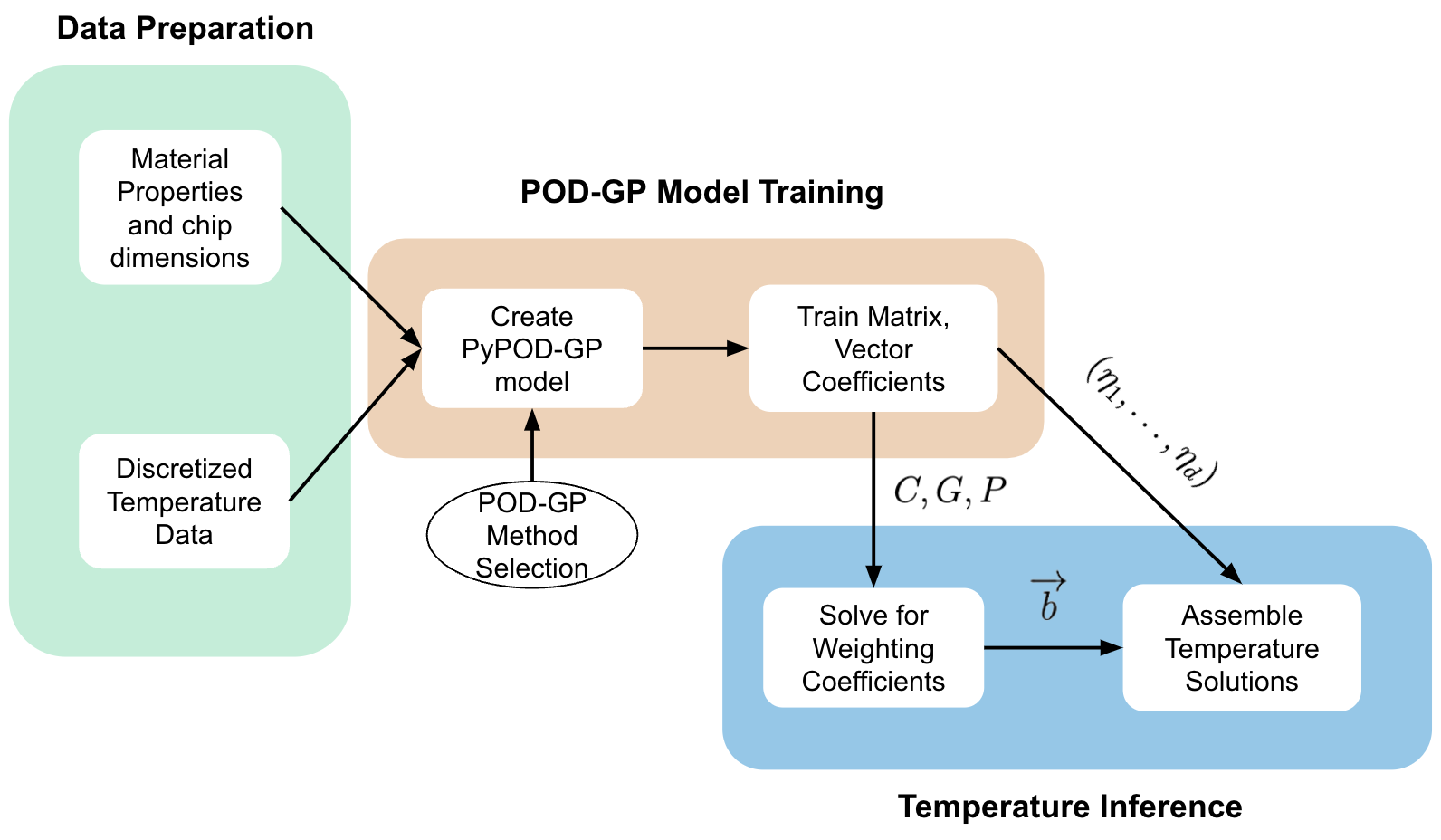}}
    \caption{Flow Chart of the Pipeline for PyPOD-GP.}
    \label{fig:pipeline}
  \end{center}
  \vspace{-40pt}
\end{wrapfigure}

\textbf{The PyPOD-GP Pipeline.}
The PyPOD-GP pipeline consists of three parts: (1) Data Preparation, (2) Model Training, and (3) Temperature Inference, as outlined in ~\cref{fig:pipeline}. The library supports training a single POD-GP model but supports inference on multiple POD-GP models. The users specify parameters for the sampled temperature data, such as chip dimensions and material properties, and provide the path to temperature data in HDF5 format. The pipeline is then executed using a \texttt{PyPOD\_GP} object. The \texttt{train} function trains the model, saving the POD modes and returning the $C$, $G$ matrices and the $\vec{P}$ vector. When training for multiple models, these values are returned as lists. The \texttt{infer} function solves the ODEs using the trained coefficients (in parallel for multiple models), and the \texttt{predict\_thermal} function generates the temperature solutions. 

\subsection{Software Functionality}
In this section, we detail the steps for thermal simulation. The entire POD-GP process is managed by a \texttt{PyPOD\_GP} object.

\textbf{Data Processing.} After initializing the \texttt{PyPOD\_GP} object with appropriate parameters and invoking the \texttt{train} function, the model processes discretized temperature data by creating mappings between cells, vertex coordinates, and their corresponding degrees of freedom (DoF). Cell properties, such as the determinant of the Jacobian transformation between global and local cell views ($\det(J_c)$), are pre-computed, and boundary surface cells are identified. 

\textbf{Model Training.} After data processing, the model calculates the coefficients described in Sec.~\ref{arch}. Integration is performed using Gaussian quadrature, representing the integration as a weighted sum of function values at each cell scaled by $|\det(J_c)|$. Function values are interpolated at $k$ quadrature points (user-specified quadrature degree) via the \texttt{get\_interpolation} and \texttt{get\_quad\_points} methods (see~\cite{larson2013fem}). The model assumes the data is formatted as a tetrahedron mesh as the chip can be seen as a rectangular prism. During POD-GP execution, the correlation matrix is computed using the \texttt{calc\_A} function. The \texttt{get\_modes} function extracts the top $d$ normalized eigenvectors $\eta_i$, mapped to domain eigenfunctions via cell vertex mappings, and saved in matrix form $U$. The \texttt{calc\_C}, \texttt{calc\_G}, and \texttt{calc\_P} functions then compute $C$, $G$, and $\vec{P}$, respectively. All computations are performed on the GPU, leveraging vectorized operations for efficiency by copying tensors from the CPU to the GPU and ensuring domain-wide vector manipulations.

\textbf{Model Inference.} After training, the \texttt{infer} function solves the ODEs to compute weighting coefficients $\vec{b}$ for each time step. A \texttt{POD\_ODE\_Solver} object solves the ODEs using a fourth-order Runge–Kutta method, returning the coefficients in matrix $B$, where rows index time and columns index space. The \texttt{predict\_thermal} function generates predicted temperatures across the domain by computing $B[i]^T U$, where $B[i]$ is the $i$-th row.

\section{Illustrative example}
This section demonstrates chip-level thermal simulation on an NVIDIA Tesla Volta GV100 GPU with 13,440 cores. We evaluate the efficiency by comparing its runtime to previous CPU-based implementations. We also report the error across the device's heating layer to validate the accuracy of our library.

\subsection{Local Ensemble POD}\label{enpod}
\begin{wrapfigure}{R}{0.40\textwidth}
\centering
\captionsetup{justification=centering}
\includegraphics[width=0.40\textwidth]{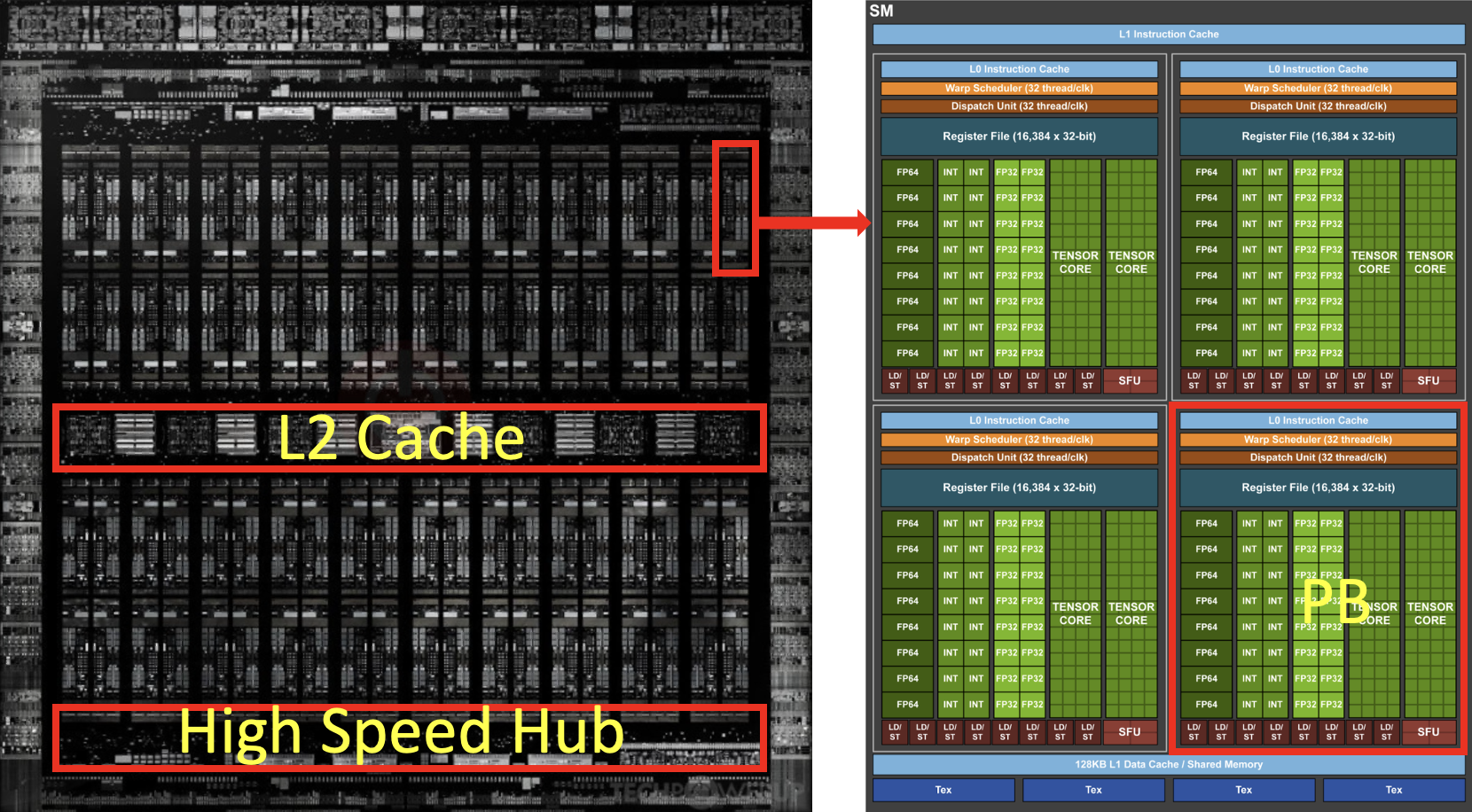}
    \caption{GPU Floor Plan}
    \label{floor_plan}
    \vspace{-10pt}
\end{wrapfigure}
A localized method similar to~\cite{lin2024ensemble} was used in this example. The chip was divided into smaller units called heat source blocks (HSBs). Since the temperature induced by an HSB becomes negligible beyond a few thermal lengths, data was sampled from a truncated domain around each HSB. Separate POD-GP processes were trained for each HSB, and the results were assembled to simulate the chip's overall thermal behavior. For the NVIDIA Tesla Volta GV100 GPU, there are 80 streaming multiprocessors (SMs). Each SM comprises four texture units and four processing blocks (PBs). Fig.~\ref{floor_plan} shows a floor plan of the chip. A total of 404 HSBs were selected, each representing individual PBs, groups of four texture units within an SM, L2 cache, high-speed hub, or one of the two memory interfaces. As the chip's has many repetitive units, a single POD-GP model can be trained to represent multiple identical HSBs. As a result, only 16 POD-GP models were needed to simulate the thermal effects of all 404 HSBs. For each model, data was sampled over 2000 time steps with a discretization of over 4 million cells per HSB.
\subsection{Runtime Analysis}
To verify the faster runtime of our implementation compared to FEniCS- and PETSc-based methods, PyPOD-GP was tested using the experimental setup described above, running on an NVIDIA Tesla Volta GV100 48GB GPU. The baseline POD-GP process was executed on an Intel(R) Xeon(R) Gold 6130 CPU with 20 MPI processes, utilizing FEniCS for training and PETSc for inference, as implemented in~\cite{lin2023podtherm}. Although only 16 POD-GP models were trained, the total computation time for 404 HSBs was estimated by scaling the runtime for HSBs of the same size. Comparison for training runtime is shown in ~\cref{train_time} and  for inference in ~\cref{infer_time}. PyPOD-GP achieves over \textbf{23.4$\times$ speedup} during training and over \textbf{10$\times$} during inference with three or more modes, highlighting its efficiency. Both the GPU and CPU used for the experiment were released in 2017, ensuring a fair runtime comparison.
\begin{table}[]
    \begin{subtable}{.5\linewidth}
      \centering
        \resizebox{\textwidth}{!}{%
\begin{tabular}{@{}llll@{}}
\toprule
\textbf{Task}       & \textbf{Method}         & \textbf{Runtime(s)}       & \textbf{Improvement} \\ \midrule
A Matrix            & FEniCS(CPU)             & $5.75\times 10^7$         & -                        \\

A Matrix            & \textbf{PyPOD-GP} & $\mathbf{2.46\times10^6}$ & $\mathbf{23.4}$\textbf{X}         \\
C Matrix (50 Modes) & FEniCS(CPU)             & $1.39\times 10^4$         & -                        \\

C Matrix (50 Modes) & \textbf{PyPOD-GP}         & $\mathbf{7.82\times10}$   & $\mathbf{177.7}$\textbf{X}          \\
G Matrix (50 Modes) & FEniCS(CPU)             & $1.37\times 10^4$         & -                        \\

G Matrix (50 Modes) & \textbf{PyPOD-GP}        & $\mathbf{1.29\times10^2}$ & $\mathbf{106.2}$\textbf{X}          \\
P Vector (20 Modes) & FEniCS(CPU)             & $7.68\times 10^3$         & -                        \\

P Vector (20 Modes) & \textbf{PyPOD-GP} & $\mathbf{2.18\times10^2}$ & $\mathbf{35.2}$\textbf{X}           \\ \bottomrule
\end{tabular}%
}\caption{Training time}\label{train_time}
    \end{subtable} 
    \begin{subtable}{.5\linewidth}
      \centering
        \resizebox{0.8\textwidth}{!}{\begin{tabular}{@{}llll@{}}
\toprule
\textbf{\#Modes}         & \textbf{Method}         & \textbf{Runtime(s)} & \textbf{Improvement}                   \\ \midrule
2 Modes  & PETSc(CPU)              & $78.6$              & -                                          \\

2 Modes  & \textbf{PyPOD-GP}          & $\mathbf{8.9}$      & $\mathbf{8.7}$\textbf{X} \\
3 Modes  & PETSc(CPU)              & $92.9$              & -                                          \\

3 Modes  & \textbf{PyPOD-GP}          & $\mathbf{9.2}$      & $\mathbf{10.1}$\textbf{X} \\
4 Modes  & PETSc(CPU)              & $93.1$              & -                                          \\
4 Modes  & \textbf{PyPOD-GP}          & $\mathbf{8.9}$      & $\mathbf{10.5}$\textbf{X} \\
5 Modes & PETSc(CPU)              & $97.0$              & -                                          \\

5 Modes & \textbf{PyPOD-GP}          & $\mathbf{9.6}$      & $\mathbf{10.1}$\textbf{X} \\
 \bottomrule
\end{tabular}}
\caption{Inference time}\label{infer_time}
    \end{subtable}
    \caption{Runtime comparison for PyPOD-GP and baselines on 404 HSBs, in seconds}
    \label{runtime}
\end{table}
\begin{wrapfigure}{R}{0.30\textwidth}
\vspace{-10pt}
\centering
\captionsetup{justification=centering}
\includegraphics[width=0.30\textwidth]{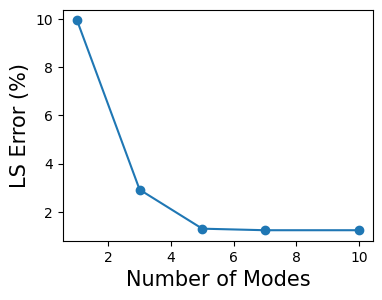}
    \caption{Error at device layer by number of modes from PyPOD-GP}
    \label{ls_error}
    \vspace{-20pt}
\end{wrapfigure}
\subsection{Accuracy Evaluation}
To validate the accuracy of PyPOD-GP, thermal simulation results on the chip's top layer (device layer) were compared to the sampled temperature data. The least squares (LS) error was computed as \begin{equation}
    \mathrm{Err} = \sqrt{\frac{\sum_{i=1}^{N_t} \int_\Omega e^2(\vec{r}, t_i) \, d\Omega}{\sum_{i=1}^{N_t} \int_\Omega \left[T(\vec{r}, t_i) - T_{\mathrm{amb}}\right]^2 \, d\Omega}},
\end{equation}
where $N_t$ is the number of time steps, $T(\vec{r}, t_i)$ is the sampled temperature at time step $i$, $T_\mathrm{amb}$ is the ambient temperature, and $e^2(\vec{r}, t_i)$ is the squared temperature difference between PyPOD-GP and the sampled temperature at time $i$. The results, shown in ~\cref{ls_error}, indicate that using 5 or more POD modes achieves less than \textbf{2\%} LS error across the device layer, with the error reducing to \textbf{1.2\%} when 7 or more modes are used.
\section{Impact}
As discussed in the introduction, effective cooling of modern chips is critical for device performance and lifespan. Modern dynamic cooling systems require thermal monitoring that is accurate, high-resolution, and efficient. While POD-GP offers an efficient and accurate solution, previous CPU-based FEM implementations suffered from significant runtime bottlenecks. Our GPU-optimized implementation, PyPOD-GP, achieves remarkable speedups, representing a significant advancement toward real-time thermal monitoring. The speedup provided by PyPOD-GP offers several practical benefits. Faster training enables the use of finer meshes for higher quality sampled temperature data, which leads to notably improved accuracy~\cite{lin2023podtherm}. Additionally, the speedup makes thermal prediction across multiple chips more feasible, such as during multi-GPU operations for training multi-head large language models~\cite{vaswani2017attention}. Finally, PyPOD-GP contributes to the research community by offering a faster tool that enables more advanced thermal simulation studies and a framework for applying POD methods to other physics problems. 
\section{Conclusions}
We present PyPOD-GP, a library implementing POD-GP methods and variants with GPU-optimized tensor operations. Compared to previous CPU-based FEM implementations using FEniCS and PETSc, PyPOD-GP achieves significant speedups in both training and inference. This enables more practical use of the POD method for real-time thermal management, particularly with higher-quality training data or multi-device predictions.

\section*{Acknowledgments}
\label{}
This project is supported by National Science Foundation under Grant No. OAC-2244049.
 \bibliographystyle{elsarticle-num} 
 \bibliography{main}

\end{document}